\documentclass[a4paper]{IEEEtran}
\usepackage{etex}
\usepackage[utf8]{luainputenc}
\usepackage{textcomp}
\DeclareUnicodeCharacter{2212}{\textminus}
\DeclareUnicodeCharacter{2011}{\mbox{-}}
\usepackage[noadjust]{cite}
\usepackage{amsmath}
\usepackage{amssymb}
\usepackage{mathtools}
\usepackage[hidelinks]{hyperref}
\usepackage{units}
\usepackage{microtype}
\usepackage{multirow}
\usepackage{booktabs}
\usepackage{tabu}
\usepackage{tikz}
\usetikzlibrary{matrix}
\usepackage{rotating}
\usepackage{subfigure}

\usepackage{setspace,tikz}
\usetikzlibrary{calc}

\newdimen{\Offset}
\newcommand{\tikzmark}[2]{%
    \settowidth{\Offset}{#2}%
    \tikz[overlay,remember picture,baseline] \node [inner sep=0pt,anchor=base] (#1) {\phantom{#2}};#2%
}

\newcommand*{\ArcDistance}{0.5cm}%
        \newcommand*{\ShortenBegin}{0pt}%
        \newcommand*{\ShortenEnd}{0pt}%
        \newcommand*{\ArcVector}{\ArcDistance}%

\usepackage{array}
\usepackage{ragged2e}
\newcolumntype{P}[1]{>{\RaggedRight\hspace{0pt}}m{#1}}

\usepackage{xspace}
\usepackage{scalerel}

\newcommand{\forkortning}[1]{\relax\ifmmode \scaleobj{0.888}{\mathsf{#1}} \else \textsc{\MakeLowercase{#1}}\xspace\fi}

\newcommand{\SINR}{\forkortning{SINR}}
\newcommand{\PAR}{\forkortning{PAR}}

\newcommand{\ACLR}{\forkortning{ACLR}}
\newcommand{\MIMO}{\textsc{mimo}\xspace}
\newcommand{\OFDM}{\textsc{ofdm}\xspace}

\newcommand{\Expectation}[1]{\ensuremath{\operatorname{\raisebox{-0.5pt}{$\mathsf{E}$}}\!\left[\,#1\,\right]}}
\DeclareMathOperator*{\argmin}{arg\,min}
\DeclareMathOperator*{\argmax}{arg\,max}
\AtBeginDocument{

}
\newcommand{\ordo}{\ensuremath{\operatorname{\mathcal{O}}}}

\newcommand{\estfreqchannel}{\ensuremath{\smash{\mathbf{\hat{\tilde{H}}}}}\rule{0pt}{0.8em}}

\newcommand{\varr}{\includegraphics[height=1ex]{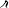}}
\newcommand{\vard}{\hskip-0.1em\includegraphics[height=2ex]{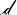}\hskip-0.3em}

\allowdisplaybreaks

\newtheorem{proposition}{Proposition}

\title{Waveforms for the Massive MIMO Downlink: Amplifier Efficiency, Distortion and Performance}
\author{Christopher Mollén, Erik G. Larsson and Thomas Eriksson %\rlap{\quad{\tiny Version: 2.\arabic{run}}} %
\thanks{C. Mollén and E. Larsson are with the Communication Systems Division, Dept.\ of Electrical Eng.\ (\textsc{isy}), Linköping University, Linköping, Sweden, and T. Eriksson is with the Dept.\ of Signals and Systems, Chalmers University of Technology, Gothenburg, Sweden.} %
\thanks{The research leading to these results has received funding from the European Union Seventh Framework Programme under grant agreement number ICT-619086 (\textsc{mammoet}), the Swedish Research Council (Vetenskapsrådet) and \textsc{elliit}.} %
\thanks{Parts of this work were presented at the European Wireless conference 2014 \cite{mollen2014impact}.}}

\begin{document}

\maketitle

\begin{abstract}
In massive \MIMO, most precoders result in downlink signals that suffer from high \PAR, independently of modulation order and whether single-carrier or \textsc{ofdm} transmission is used.  The high \PAR lowers the power efficiency of the base station amplifiers.  To increase power efficiency, low-\PAR precoders have been proposed.  In this article, we compare different transmission schemes for massive \MIMO in terms of the power consumed by the amplifiers.  It is found that (i) \OFDM and single-carrier transmission have the same performance over a hardened massive \MIMO channel and (ii) when the higher amplifier power efficiency of low-\PAR precoding is taken into account, conventional and low-\PAR precoders lead to approximately the same power consumption.  Since downlink signals with low \PAR allow for simpler and cheaper hardware, than signals with high \PAR, therefore, the results suggest that low-\PAR precoding with either single-carrier or \OFDM transmission should be used in a massive \MIMO base station.
\end{abstract}


\section{Introduction}
\IEEEPARstart{W}{ireless} massive \MIMO systems, initially conceived in \cite{marzetta2010noncooperative} and popularly described in \cite{6690}, simultaneously serve tens of users with base stations equipped with tens or hundreds of antennas using multiuser precoding.  Compared to classical multiuser \MIMO, order-of-magnitude improvements are obtained in spectral and energy efficiency \cite{ngo2013energy, yang2013performance}.  For these reasons, massive \MIMO is expected to be a key component of future wireless communications infrastructure \cite{6690, hoydis2013making}.

This work compares different multiuser precoding schemes for the massive \MIMO downlink.  Under a \emph{total radiated} power constraint, optimal multiuser \MIMO precoding is a rather well-understood topic, see, e.g., \cite{caire2003achievable, goldsmith2003capacity}, as is linear (and necessarily suboptimal) precoding, see, e.g., \cite{joham2005linear} and the survey \cite{6832894}.  It is also known that, for massive \MIMO specifically, linear precoding is close to optimal under a total-radiated power constraint \cite{yang2013performance}.  There are also numerous results on precoding under \emph{per-antenna} power constraints \cite{vu2011miso, vu2011mimo, yu2007transmitter}.

\begin{figure}
\centering
\includegraphics[width=229pt]{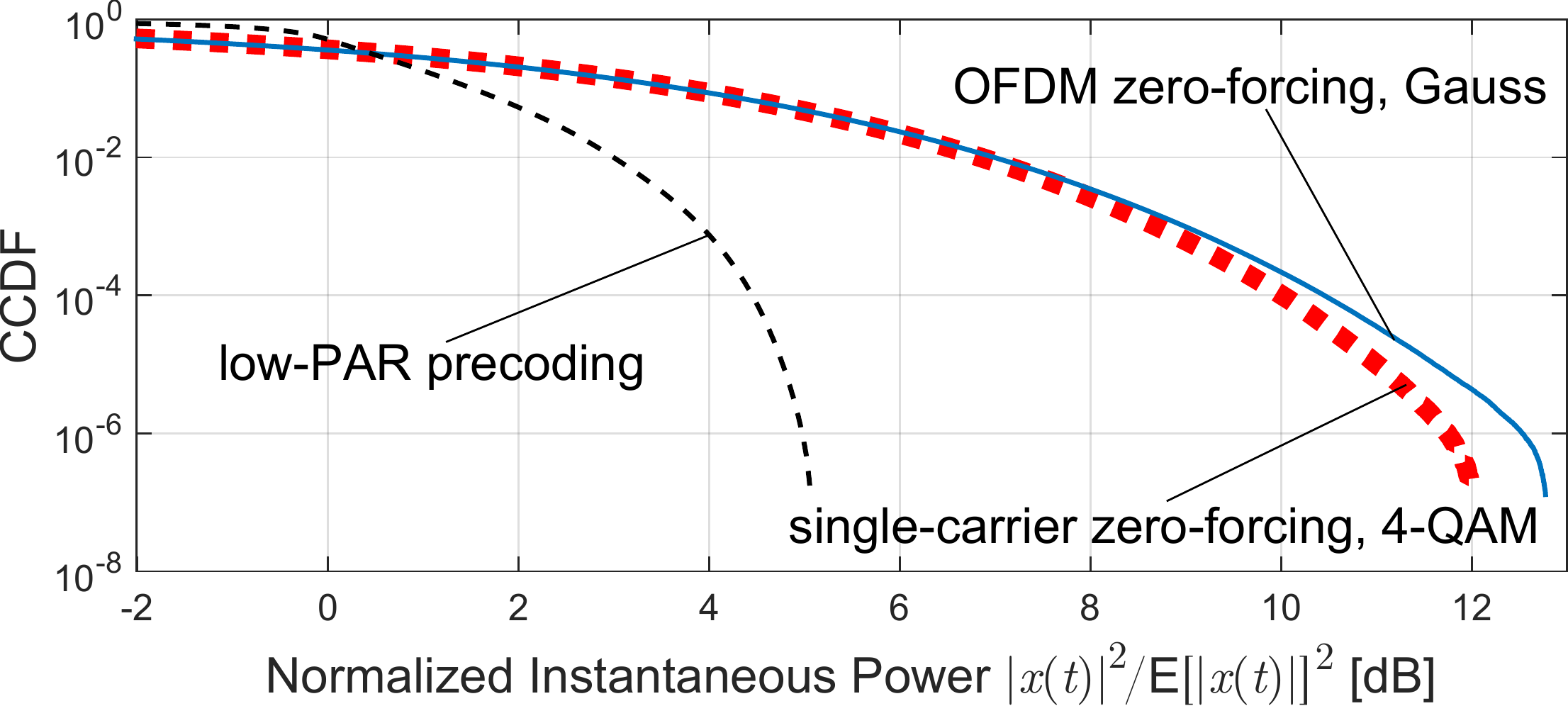}
\caption{The complimentary cumulative distribution of the amplitudes of different massive \MIMO downlink signals that have been pulse shaped by a root-raised cosine filter, roll-off 0.22.  Single-carrier and \OFDM transmission have very similar distributions for the linear precoders described in Section~\ref{sec:linear_precoders} for any modulation order, cf.\ \cite{6457368}.  The low-\PAR precoding scheme is described in \cite{6565529} and in Section~\ref{sec:DTCE}.  The system has 100 base station antennas, 10 single-antenna users and the channel is i.i.d.\ Rayleigh fading with 4 taps.}
\label{fig:amp_dist}
\end{figure}

In practice, massive \MIMO precoders optimized subject to a total radiated power constraint yield transmit signals with high peak-to-average ratio (\PAR), regardless of whether single-carrier or \OFDM transmission and whether a low or a high modulation order is used, see Figure~\ref{fig:amp_dist}.  To avoid heavy signal distortion and out-of-band radiation, transmission of such signals requires that the power amplifiers are backed off and operated at a point, where their transfer characteristics are sufficiently linear \cite{hemesi2013analytical}.  The higher the signal \PAR is, the more backoff is needed; and the higher the backoff is, the lower the power efficiency of the amplifier will be.  Against this background, precoders that yield signals with low \PAR would be desirable.

The possibility to perform low-\PAR precoding is a unique opportunity offered by the \emph{massive} \MIMO channel---signal peaks can be reduced because the massive \MIMO downlink channel has a large nullspace and any additional signal transmitted in the nullspace does not affect what the users receive.  In particular, \PAR-reducing signals from the channel nullspace can be added to the downlink signal so that the emitted signals have low \PAR \cite{6451071,6690}.  A few low-\PAR precoders for massive \MIMO have been proposed in the literature \cite{6565529, 6451071, 6415400}.  In \cite{6451071}, the discrete-time downlink signals were constrained to have constant envelopes.\footnote{Note that the precoder in \cite{6451071} can transmit symbols from any general input constellation with single-carrier transmission and \OFDM---the received signals do not have to have constant envelopes, only the downlink signal emitted from each base station antenna has constant envelope.}  There it was estimated that, in typical massive \MIMO scenarios, \unit[1--2]{dB} extra radiated power is required to achieve the same sum-rate as without an envelope constraint.  While some extra radiated power is required by low-\PAR precoders, it was argued in \cite{6451071} that the overall power consumption still should decrease due to the increased amplifier efficiency.  

Another unique feature of the massive \MIMO downlink channel is that certain types of hardware-induced distortion tend to average out when observed at the receivers \cite{bjornson2013massive}.  Our study confirms that the variance of the in-band distortion caused by nonlinear base station amplifiers does decrease with the number of base station antennas.

The objective of the paper at hand is to more accurately quantify the benefits of low-\PAR precoding for the massive \MIMO downlink, taking into account in-band distortion and out-of-band radiation stemming from amplifier nonlinearities and imperfect channel state information due to pilot-based channel estimation.  The difference between \OFDM and single-carrier transmission is also investigated.  The main technical contribution of the paper is a comprehensive end-to-end modelling of massive \MIMO downlink transmission, which is treated in continuous time in order to capture the effects of nonlinear amplification, the associated capacity bound, and the estimations of the power consumption for relevant amplifier models.  All conclusions are summarized in Section~\ref{sec:conclusions}.

We stress that the effect of amplifier nonlinearities on wireless signals have also been studied by others \cite{costa1999impact, hemesi2013analytical}, and for \MIMO specifically in \cite{qi2010analysis}.  In relation to this literature, the novel aspects of our work include: (i) a specific focus on the massive \MIMO downlink channel, which facilitates low-\PAR precoding; (ii) a classification and comparison of precoders commonly considered for massive \MIMO; (iii) an estimate of the amplifier power consumption of low-\PAR precoding in comparison to that of other standard precoders.

\section{System Model}
\begin{figure}\label{fig:system_model}
	\centering %\addtocounter{figure}{1}
	\subfigure[The continuous-time model of the downlink.]{\label{fig:cont_system_model}\includegraphics[width=\linewidth]{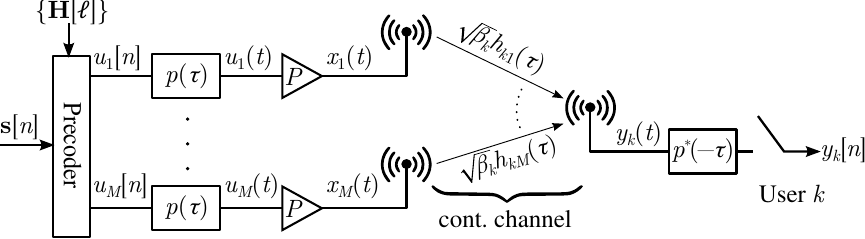}\hspace*{3em}}
	\subfigure[The equivalent discrete-time model of the downlink.]{\label{fig:disc_system_model}\includegraphics[width=\linewidth]{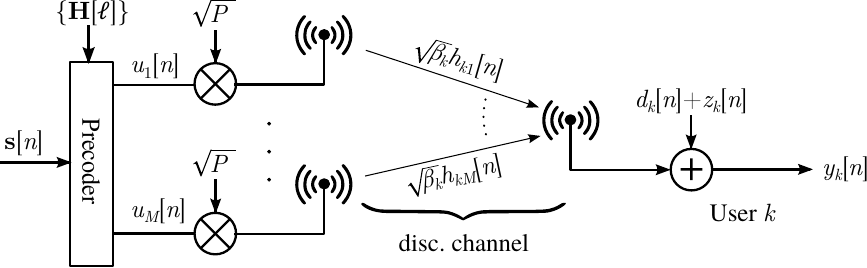}\hspace*{3em}}
	\caption{The downlink of a massive \MIMO system.}
\end{figure}

The downlink shown in Figure~\ref{fig:cont_system_model} is studied.  The base station is equipped with $M$ antennas and it serves $K$ single-antenna users over a frequency-selective channel.  All signals are modeled in complex baseband.

We let $s_k[n]$ be the $n$\mbox{-}th symbol that is to be transmitted to user $k$ and collectively denote all the $n$\mbox{-}th symbols by $\mathbf{s}[n] \triangleq (s_1[n], \ldots, s_K[n])$.  The base station precodes the symbols to produce the discrete-time signals $\{u_m[n]\}$, where $u_m[n]$ is the precoded signal of antenna $m$.  These signals are scaled such that
\begin{align}\label{eq:power_cosntraint}
	\sum_{m = 1}^M \Expectation{|u_m[n]|^2} = 1, \quad\forall n
\end{align}
and pulse shaped by a filter with impulse response $p(\tau)$ into the continuous-time transmit signals
\begin{align}\label{eq:psf_tx_signal}
	u_m(t) \triangleq \sum_{n}u_m[n]p(t - nT),
\end{align}
where $T$ is the symbol period.  After pulse shape filtering, the transmit signal $u_m(t)$ has a bandwidth smaller or equal to the bandwidth $B$ of the pulse $p(\tau)$.  The bandwidth $B$ is the width of the interval, over which the spectrum of $p(\tau)$ is non-zero.  For example, if a root-raised cosine filter of period $T$ with roll-off $\sigma $ were used, then $BT = 1 + \sigma$.  

The continuous-time signal $u_m(t)$ is then amplified to transmit power by an amplifier that, in general, is nonlinear.  The amplified signal is given by
\begin{align}\label{eq:IO_amp}
x_m(t) = g\big(\tfrac{|u_m(t)|}{\sqrt{b}}\big)e^{j(\arg u_m(t) + \Phi(|u_m(t)|/\sqrt{b}))},
\end{align}
where $g(|u_m(t)|)$ is the \textsc{am}-\textsc{am} conversion and $\Phi(|u_m(t)|)$ the \textsc{am}-\textsc{pm} conversion, see for example \cite{6457368}.  For now, the conversions $g(u)$ and $\Phi(u)$ are generic functions.  Later in our analysis however, appropriate assumptions will be made to specify them.  The factor $b$ is the backoff that has to be done to avoid nonlinear amplification and distortion.  By backing off the signal power to a suitable operating point, the signal amplitude will stay in a region with sufficiently linear amplification most of the time, see \cite{6457368}.  In this article, all backoffs are given in \unit{dB} relative to the backoff of the 1-dB compression point---the point, where the output signal is \unit[1]{dB} weaker than what it would have been if the amplification were perfectly linear.  The signals are amplified so that
\begin{align}\label{eq:power_constraint}
	\lim_{t_0 \to \infty} \sum_{m = 1}^{M} \mathsf{E}\Bigg[ \frac{1}{t_0} \int\limits_{\mathclap{-t_0/2}}^{\mathclap{t_0/2}} |x_m(t)|^2 \mathrm{d}t\Bigg] = P,
\end{align}
where $P$ is the transmitted power of the base station.  

\begin{figure}
	\centering
	\mbox{\hspace{4.7em}\includegraphics{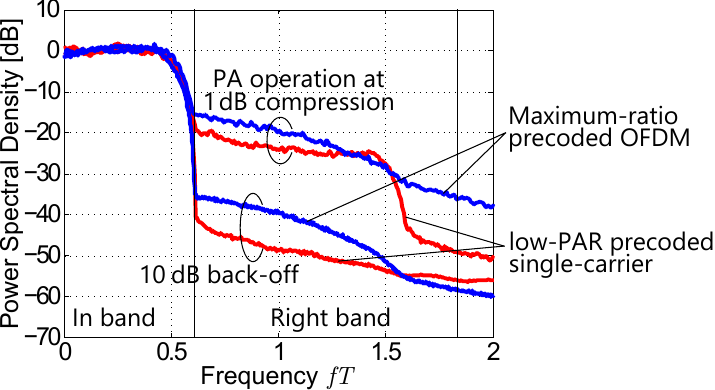}}
	\caption{The power spectral densities after amplification of two signal types with the \textsc{pa} operating at the \unit[1]{dB} compression point (upper two curves) and well below saturation (lower two curves).  The signals are from the system later described in Table~\ref{tab:sim_param}, where the bandwidth of the pulse $p(\tau)$ is $BT = 1.22$.}
	\label{fig:PSD}
\end{figure}

The nonlinear relation in \eqref{eq:IO_amp} generally widens the spectrum of the amplified signal, i.e.\ its signal energy is no longer confined to the bandwidth $B$ of the pulse $p(\tau)$.  The energy outside the ideal bandwidth is called \emph{out-of-band radiation} and is quantified by the \emph{Adjacent Channel Leakage Ratio} (\ACLR), which is defined in terms of the power $P_{[-B/2,B/2]}$ of $x_m(t)$ in the useful band and the powers $P_{[-3B/2,-B/2]}$, $P_{[B/2,3B/2]}$ in the immediately adjacent bands:  
\begin{align}
\ACLR \triangleq  \max \left(\frac{P_{[-3B/2,-B/2]}}{P_{[-B/2,B/2]}},\frac{P_{[B/2,3B/2]}}{P_{[-B/2,B/2]}}\right),
\end{align}
where 
\begin{align}
P_\mathcal{B} \triangleq  \int_{f \in \mathcal{B}} S_x(f) \mathrm{d}f.
\end{align}
and $S_x(f)$ is the power spectral density of $x_m(t)$.  In Figure~\ref{fig:PSD}, four power spectral densities of different amplified signals are shown to illustrate the out-of-band radiation.  Half the in-band spectrum is shown together with the whole right band.  

This signal is broadcast over the channel, whose small-scale fading impulse response from antenna $m$ to user $k$ is $h_{km}(\tau)$ and large-scale fading coefficient to user $k$ is $\beta_k$.  Specifically, user $k$ receives the signal 
\begin{align}
	y_k(t) = \sqrt{\beta_k} \sum_{m=1}^{M} \big(h_{km}(\tau) \star x_m(\tau)\big)(t) + z_k(t), 
\end{align}
where $z_k(t)$ is a stationary white Gaussian stochastic process with spectral height $N_0$ that models the thermal noise of the user equipment.  The received signal is passed through a filter matched to the pulse $p(\tau)$ and sampled to produce the discrete-time received signal
\begin{align}\label{eq:rx_sampling}
	y_k[n] \triangleq \big(p^*(-\tau) \star y_k(\tau)\big)(nT).
\end{align}

In analyzing this system, we will look into an equivalent discrete-time system, see Figure~\ref{fig:disc_system_model}.  In order to do that, the distortion produced by the nonlinear amplifier has to be treated separately, since the nonlinearity widens the spectrum and is not accurately described by symbol-rate sampling.  The small-scale fading coefficients of the discrete-time impulse response of the channel, including the pulse-shaping and matched filter, between antenna $m$ and user $k$ are denoted
\begin{align}
	h_{km}[\ell] \triangleq T \big(p(\tau) \star h_{km}(\tau) \star p^*(-\tau)\big)(\ell T).
\end{align}
For these channel coefficients, we assume that 
\begin{align}
\Expectation{h_{km}^*[\ell] h_{km}[\ell']} &= 0, \quad \forall \ell \neq \ell',\\
\sum_{\mathclap{\ell = 0}}^{\mathclap{L - 1}}  \Expectation{|h_{km}[\ell]|^2} &= 1, \quad \forall k,m,
\end{align}
and that $h_{km}[\ell]$ is zero for integers $\ell \notin [0, L{-}1]$, where $L$ is the number of channel taps.

The $n$\mbox{-}th received sample at user $k$ is then given by
\begin{align}\label{eq:disc_rx_signal}
	y_k[n] = \sqrt{P \beta_k} \Big(\sum_{\mathclap{m=1}}^M\! \sum_{\mathclap{\ell=0}}^{L-1}\! h_{km}[\ell] u_m[n{-}\ell] + d_k[n]\Big) + z_k[n],
\end{align}
where the noise sample $z_k[n] \triangleq \big(p^*(-\tau) \star z_k(\tau)\big)(nT)$.  To make $z_k[n] \sim \mathcal{CN}(0, N_0 / T)$ i.i.d., it is assumed that $p(\tau)$ is a root-Nyqvist pulse of period $T$ and signal energy $1/T$.  The term $d_k[n]$ describes the \emph{in-band distortion}---the part of the distortion that can be seen in the received samples $y_k[n]$---caused by the nonlinear amplification of the transmit signal.  It is given by $d_k[n] \triangleq$
\begin{align}
	\frac{1}{\sqrt{P}} \!\sum_{\mathclap{m=1}}^M\!\! \Big(\!\big(x_m(t) {-} \sqrt{\!P} u_{m}\!(t)\big) \star h_{km}(\tau) \star p^*\!(\!-\tau)\!\Big)\!(nT).
\end{align}

By introducing the following vectors 
\begin{align*}
\mathbf{u}[n] &\triangleq (u_1[n],\ldots,u_M[n])^\mathsf{T} & \mathbf{y}[n] &\triangleq (y_1[n], \ldots, y_K[n])^\mathsf{T}\\
\mathbf{d}[n] &\triangleq (d_1[n], \ldots, d_K[n])^\mathsf{T} & \mathbf{z}[n] &\triangleq (z_1[n], \ldots, z_K[n])^\mathsf{T}
\end{align*}
and matrices
\begin{align}\label{eq:def_channel_matrix}
	\mathbf{H}[\ell] &{\triangleq}\! \begin{pmatrix}
h_{11}[\ell] & \mathclap{\cdots} & h_{1M}[\ell]\\
\vdots & \mathclap{\ddots} & \vdots\\
h_{K1}[\ell] & \mathclap{\cdots} & h_{KM}[\ell]\\
\end{pmatrix} & \mathbf{B} &{\triangleq}\operatorname{diag}(\beta_1, \ldots, \beta_K),
\end{align}
the received signals can be written as
\begin{align}\label{eq:time_RX_signal}
	\mathbf{y}[n] = \sqrt{P}\mathbf{B}^\frac{1}{2} \Big(\sum_{\mathclap{\ell = 0}}^{\mathclap{L - 1}} \mathbf{H}[\ell] \mathbf{u}[n - \ell] + \mathbf{d}[n]\Big) + \mathbf{z}[n].
\end{align}

If the transmission were done in a block of $N$ symbols per user, and a \emph{cyclic prefix} were used in front of the blocks, i.e.\
\begin{align}\label{eq:cyclic_prefix}
\mathbf{u}[n] = \mathbf{u}[N+n], \quad\text{for }n=-L,\ldots,-1,
\end{align}
where $n = 0$ is the time instant when the first symbol is received at the users, then the received signal in \eqref{eq:time_RX_signal} is easily given in the frequency domain.  If the discrete Fourier transforms of the transmit signals, received signals and channel are denoted by
\begin{align}
\mathbf{\tilde{u}}[\nu] &\triangleq \sum_{\mathclap{n = 0}}^{\mathclap{N - 1}} e^{-j2\pi n \nu / N} \mathbf{u}[n], \label{eq:freq_tx_signal}\\
\mathbf{\tilde{y}}[\nu] &\triangleq \sum_{\mathclap{n = 0}}^{\mathclap{N - 1}}e^{-j 2 \pi n \nu / N} \mathbf{y}[n], \label{eq:freq_rx_signal}\\
\mathbf{\tilde{H}}[\nu] &\triangleq \sum_{\mathclap{\ell = 0}}^{\mathclap{L - 1}}e^{-j2\pi \ell \nu / N}\mathbf{H}[\ell],\label{eq:freq_channel}
\end{align}
then the received signal at frequency index $\nu$ is given by
\begin{align}
\mathbf{\tilde{y}}[\nu] = \sqrt{P} \mathbf{B}^\frac{1}{2} \big(\mathbf{\tilde{H}}[\nu]\mathbf{\tilde{u}}[\nu] + \mathbf{\tilde{d}}[\nu] \big) + \mathbf{\tilde{z}}[\nu],
\end{align}
where $\mathbf{\tilde{d}}[\nu]$ describes the in-band distortion caused by the nonlinear amplification, $\mathbf{\tilde{z}}[\nu] \sim \mathcal{CN}(0, \tfrac{N_0}{T}\mathbf{I}_K)$ and $\mathbf{I}_K$ is the $K\!$‑dim\-en\-sion\-al identity matrix.  The frequency-domain notation in \eqref{eq:freq_tx_signal}--\eqref{eq:freq_channel} will be useful when we later consider \OFDM-based transmission methods.  

In this paper, we limit ourselves to look at block transmission with a cyclic prefix \eqref{eq:cyclic_prefix}.  To require a cyclic prefix simplifies our exposition and does not limit its generality much.  A prefix is present in almost all modern digital transmission schemes, as a guard interval or as a delimiter between blocks.  A prefix that is correlated with the symbols is arguably a waste of spectral resources.  However, by letting $N$ be much greater than $L$, this waste can be made arbitrarily small. 

\section{Downlink Transmission}
In the downlink, a precoder chooses, based on the channel state information available at the base station, transmit signals such that the users receive the symbols intended for them.  The symbols to be transmitted fulfil
\begin{align}
	\Expectation{|s_k[n]|^2} &= \xi_k, \quad\forall n, k,\label{eq:unit_energy_symbols}
\end{align}
where $\{\xi_k\}$ are power allocation coefficients that are normalized such that
\begin{align}\label{eq:power_allocation_factor_constraint}
\sum_{\mathclap{k = 1}}^K \xi_k = 1.
\end{align}

We assume that the uplink and downlink are separated in time, using so called time-division duplexing, and that each user sends an $N_\text{p}$\mbox{-}symbol long pilot sequence in the uplink that is orthogonal to the pilots of all other users.  The pilots are used by the base station to estimate the small-scale fading coefficients $\{h_{km}[\ell]\}$.  The large scale fading coefficients $\{\beta_k\}$ are assumed to be known.  Note that, to achieve orthogonality between pilots, $N_\text{p} \geq KL$.  Further, it is assumed that the channel estimates
\begin{align}\label{eq:channel_estimates}
	\hat{h}_{km}[\ell] = h_{km}[\ell] - \epsilon_{km}[\ell], \quad \forall k,m,\ell,
\end{align}
where $\epsilon_{km}[\ell]$ is the estimation error, are obtained through linear minimum-mean-square estimation, so that $\hat{h}_{km}[\ell]$ and $\epsilon_{km}[\ell]$ are uncorrelated.  In analogy with \eqref{eq:freq_channel}, we will denote the Fourier transforms of the channel estimates and the estimation error $\{\hat{\tilde{h}}_{km}[\nu]\}$ and $\{\tilde{\epsilon}_{km}[\nu]\}$ respectively.  Their variances are
\begin{align}
	\delta_k &\triangleq \sum_{\ell=0}^{\mathclap{L-1}} \Expectation{|\hat{h}_{km}[\ell] |^2} = \Expectation{|\hat{\tilde{h}}_{km}[\nu]|^2},\\
	\mathcal{E}_k &\triangleq \sum_{\ell=0}^{\mathclap{L-1}} \Expectation{|\epsilon_{km}[\ell]|^2} = \Expectation{|\tilde{\epsilon}_{km}[\nu]|^2}.
\end{align}

Note that if $\{h_{km}[\ell]\}$ are i.i.d.\ across $k$ and $m$ and if the uplink is perfectly linear, then 
\begin{align}\label{eq:if_iid}
	\delta_k = \frac{N_\text{p} \rho_\text{p} \beta_k}{1 + N_\text{p} \rho_\text{p} \beta_k}, \quad \mathcal{E}_k = \frac{1}{1 + N_\text{p} \rho_\text{p} \beta_k},
\end{align}
where $\rho_\text{p}$ is the ratio between the power used to transmit the pilots and the thermal noise variance of a base station antenna.  

\subsection{Achievable Data Rates}\label{sec:rx_architecture}
To treat single-carrier and \OFDM transmission together, let 
\begin{align}
\bar{y}_k[n] \triangleq \begin{cases}
y_k[n], &\text{if single-carrier transmission}\\
\tilde{y}_k[n], &\text{if \textsc{ofdm} transmission}
\end{cases}
\end{align}
be the $n$\mbox{-}th received sample at user $k$.  A lower bound on the capacity of the downlink channel to user $k$ is given by \cite{bjornson2013massive}
\begin{align}\label{eq:original_bound}
R_k {\triangleq} \log_2\left(1 + \SINR_k\right),
\end{align}
where the signal-to-interference-and-noise ratio is given by
\begin{align}\label{eq:original_SINR}
	\SINR_k = \frac{|\Expectation{\bar{y}_k^*[n]s_k[n]}|^2 / \xi_k}{\Expectation{|\bar{y}_k[n]|^2} - |\Expectation{\bar{y}_k^*[n] s_k[n]}|^2 / \xi_k}
\end{align}

To evaluate \eqref{eq:original_bound}, we consider the following signal:
\begin{align}
	r_k[n] = 
	\begin{cases}
		\sum_{m=1}^{M} \sum_{\ell = 0}^{L - 1} \hat{h}_{km}[\ell] u_m[n - \ell], &\text{if \textsc{sc}}\\
		\sum_{m=1}^{M} \hat{\tilde{h}}_{km}[n] \tilde{u}_m[n], &\text{if \textsc{ofdm}}
	\end{cases}.
\end{align}
and define the deterministic constant
\begin{align}\label{eq:channel_est}
g_k \triangleq \frac{1}{\sqrt{\delta_k}\xi_k} \mathsf{E}\big[s_k^*[n] r_k[n] \big].
\end{align}
Here $g_k$ is normalized by $\sqrt{\delta_k}$ so that it does not depend on the estimation error.  This normalization will later allow us to see the impact of the channel estimation error on the \SINR.  The fact that $g_k$ does not depend on $\delta_k$ is seen by expanding $g_k$ as is done in \eqref{eq:array_gain_step2} in the Appendix.  

Now the received signal can be written
\begin{align}\label{eq:rx_signal}
\bar{y}_k[n] = \sqrt{P\beta_k}\big(g_k \sqrt{\delta_k} s_k[n] {+} i_k[n] {+} e_k[n] {+} \bar{d}_k[n]\big) + \bar{z}_k[n].
\end{align}

In this sum, the first term is equal to the signal of interest, scaled by $g_k \sqrt{\delta_k}$.  The second term
\begin{align}
i_k[n] \triangleq r_k[n] - g_k \sqrt{\delta_k} s_k[n]
\end{align}
is a term comprising interference and downlink channel gain uncertainty (as in \cite{yang2013performance}).  The third term in \eqref{eq:rx_signal}
\begin{align}
	e_k[n] \triangleq
	\begin{cases}
		\sum_{m=1}^{M} \sum_{\ell = 0}^{L - 1} \epsilon_{km}[\ell] u_m[n-\ell], &\text{if \textsc{sc}}\\
		\sum_{m=1}^{M} \tilde{\epsilon}_{km}[n] \tilde{u}_m[n], &\text{if \OFDM}
	\end{cases}
\end{align}
is the error due to imperfect channel state knowledge at the base station.  The last two terms in \eqref{eq:rx_signal} are the in-band distortion that the users see because of nonlinear amplification and thermal noise respectively:
\begin{align}
\bar{d}_k[n] \triangleq
\begin{cases}
	d_k[n]\\
	\tilde{d}_k[n]
\end{cases}\quad
\bar{z}_k[n] \triangleq 
\begin{cases}
	z_k[n], &\text{if \textsc{sc}}\\
	\tilde{z}_k[n], &\text{if \OFDM}
\end{cases}.
\end{align}

The interference $i_k[n]$ is uncorrelated with $s_k[n]$ in \eqref{eq:rx_signal}, because
\begin{align}
\Expectation{s_k^*[n]i_k[n]} &= \mathsf{E}\big[s_k^*[n] (r_k[n] - g_k \delta_k^{\frac{1}{2}} s_k[n]) \big]\notag\\
&= \mathsf{E}\big[s_k^*[n] r_k[n] \big] -  g_k \delta_k^{\frac{1}{2}} \xi_k = 0.\label{eq:i_s_uncorr}
\end{align}
The in-band distortion $\bar{d}_k[n]$, on the other hand, is correlated with $s_k[n]$ and $i_k[n] + e_k[n]$.  These two correlations, we denote, similarly to \eqref{eq:channel_est},
\begin{align}
c_k &\triangleq \frac{1}{\sqrt{\delta_k} \xi_k} \Expectation{s^*_k[n] \bar{d}_k[n]},\\
\rho_k &\triangleq \frac{1}{I_k + E_k} \Expectation{(i^*_k[n] + e^*_k[n]) \bar{d}_k[n]},
\end{align}
where
\begin{align}
E_k &\triangleq \Expectation{|e_k[n]|^2},\\
I_k &\triangleq \Expectation{|i_k[n]|^2},\label{eq:interference}
\end{align}
are the channel error and interference variances.  We note that, when \eqref{eq:if_iid} holds, $E_k = \mathcal{E}_k$.
The in-band distortion can now be divided into three parts:
\begin{align}
\bar{d}_k[n] = c_k \sqrt{\delta_k} s_k[n] + \rho_k (i_k[n] + e_k[n]) + d'_k[n],
\end{align}

The first part of the in-band distortion: $d'_k[n]$ is uncorrelated to $s_k[n]$ and $i_k[n] + e_k[n]$, for the same reason $s_k[n]$ and $i_k[n]$ are uncorrelated in \eqref{eq:i_s_uncorr}.  The factor $c_k$ is thus the amount of amplitude that the nonlinear amplification “contributes” to the amplitude of the desired signal.  Usually, in a real-world system, this is a negative contribution in the sense that $|g_k + c_k| < |g_k|$.  It should therefore be seen as the amount of amplitude lost (in what is usually called \emph{clipping}).  Similarly, the other correlation $\rho_k$ is the amount of interference that is clipped by the nonlinear amplification.  Finally, we denote the variance of the in-band distortion
\begin{align}
	D_k &\triangleq \Expectation{|d'_k[n]|^2}.\label{eq:inband_dist}
\end{align}

With this new notation, the two expectations in \eqref{eq:original_SINR} can be written as follows.
\begin{flalign}
&|\Expectation{\bar{y}_k^*[n] s_k[n]}|^2 / \xi_k = \delta_k \xi_k P \beta_k |g_k + c_k|^2&&\\
&\!\Expectation{\hskip-0.19em|\bar{y}_k[n]|^{\hskip-0.1em 2}\hskip-0.19em}\hskip-0.2em {=}
P \beta_k (\delta_k \xi_k |g_k {+} c_k|^2 &&\notag\\&\hspace{3.2cm}{+} (I_k {+} E_k)|1{+}\rho_k|^2 {+} D_k) {+} \tfrac{N_0}{T}&&
\end{flalign}

This simplifies \eqref{eq:original_SINR}, which becomes
\begin{align}\label{eq:SINR}
\SINR_k = \frac{\delta_k \xi_k P \beta_k |g_k + c_k|^2}{P \hspace{-1pt} \beta_k \hspace{-1pt} \big( \hspace{-1pt} (I_k {+} E_k)|1 {+} \rho_k|^2 {+} D_k \hspace{-1pt}\big) {+} \tfrac{N_0}{T}\hspace{-1pt}}.
\end{align}

From \eqref{eq:SINR}, the two consequences of nonlinear amplification can be seen: (i) in-band distortion with variance $D_k$ and (ii) signal clipping by $c_k$, a reduction of the signal amplitude that results in a power-loss.

We also see that the variance $\delta_k$ is the fraction between the power that would have been received if the channel estimates were perfect and the actually received power.  It can thus be seen as a measure of how much power that is lost due to imperfect channel state information at the base station.  

The bound \eqref{eq:original_bound} is an achievable rate of a system that uses a given precoder and where the detector uses \eqref{eq:channel_est} as a channel estimate and treats the error terms in \eqref{eq:rx_signal} as additional uncorrelated Gaussian noise.  This detector has proven to be close to the optimal detector in environments, where the massive \MIMO channel hardens.

\subsection{Linear Precoding Schemes}
With knowledge of the channel, the base station can \emph{precode} the symbols in such a way that the gain $g_k$ is big and the interference $I_k$ small.

\subsubsection{OFDM-Transmission}
In \OFDM transmission, the precoder is defined in the frequency domain.  The time domain transmit signals are obtained from the inverse Fourier transform
\begin{align}\label{eq:tx_signal_OFDM}
	\mathbf{u}[n] \triangleq \sum_{\nu = 0}^{N - 1} e^{j 2 \pi n \nu / N} \mathbf{\tilde{u}}[\nu]
\end{align}
of the precoded signals
\begin{align}
	\mathbf{\tilde{u}}[\nu] = \mathbf{\tilde{W}}[\nu]\mathbf{s}[\nu], \quad\nu=1,\ldots,N{-}1
\end{align}
where $\mathbf{\tilde{W}}[\nu]$ is a precoding matrix for frequency $\nu$.  The precoding is linear, because the precoding matrix does not depend on the symbols, only on the channel.

To ensure that \eqref{eq:power_cosntraint} is fulfilled, it is required that
\begin{align}\label{eq:precoder_normalization}
	\Expectation{\|\mathbf{\tilde{W}}[\nu]\|_\mathsf{F}^2} = K, \quad\forall \nu.
\end{align}

\subsubsection{Single-Carrier Transmission}
The transmit signals of single-carrier transmission are given by the cyclic convolution
\begin{align}\label{eq:tx_signal_SC}
	\mathbf{u}[n] = \sum_{\ell=0}^{N-1}  \mathbf{W}[\ell]\mathbf{s}[n-\ell],
\end{align}
where the indices are taken modulo $N$.  The impulse response of the precoder is given in terms of its frequency domain counterpart:
\begin{align}\label{eq:SC_OFDM_relation}
	\mathbf{W}[\ell] \triangleq \sum_{\nu = 0}^{N - 1} e^{j 2 \pi \nu \ell / N} \mathbf{\tilde{W}}[\nu].
\end{align}

\subsubsection{Conventional Precoders}\label{sec:linear_precoders}
In this paper, three conventional precoders are studied.  They will be given as functions of the channel estimates $\{\mathbf{\hat{H}}[\ell]\}$ (a sequence of matrices defined in terms of $\{\hat{h}_{km}[\ell]\}$ in the same way as $\mathbf{H}[\ell]$ is defined in \eqref{eq:def_channel_matrix} in terms of $\{h_{km}[\ell]\}$) and its Fourier transform
\begin{align}
	\estfreqchannel[\nu] \triangleq \sum_{\ell = 0}^{N - 1} \mathbf{\hat{H}}[\ell]e^{-j2\pi \nu \ell / N}.
\end{align}
The factors $\alpha_\mathsf{x}$ used in the definitions below are normalization constants, chosen such that \eqref{eq:precoder_normalization} holds.  

\paragraph{Maximum-Ratio Precoding}
Maximum-ratio precoding is the precoder that maximizes the gain $g_k$ and the received power of the desired signal.  It is given by
\begin{align}\label{eq:MR}
\mathbf{\tilde{W}}[\nu] = 
	\alpha_\text{MR} \estfreqchannel^\mathsf{H}[\nu],\quad\text{for \textsc{mr}}.
\end{align}
While it maximizes the received power of the transmission, interference $I_k \neq 0$ is still present in the received signal.  In typical scenarios with favorable propagation, maximum-ratio precoding suppresses this interference increasingly well with higher number of base station antennas and in the limit of infinitely many antennas, the interference becomes negligible in comparison to the received power \cite{marzetta2010noncooperative}.  For maximum-ratio precoding and an i.i.d.\ Rayleigh fading channel, both with single-carrier and \OFDM transmission, the array gain and interference terms are \cite{yang2013performance}
\begin{align*}
g_k = \sqrt{M}, \quad\quad
I_k = \delta_k, \quad\quad\text{for \textsc{mr}}.
\end{align*}

Because the precoding weights of antenna $m$ only depend on the channel coefficients $\{\hat{h}_{km}[\ell]\}$ of that antenna, maximum-ratio precoding can be implemented in a distributed fashion, where the precoding is done locally at each antenna.  

Note that this definition of maximum-ratio precoding makes it equivalent to time-reversal precoding for single-carrier transmission, see for example \cite{diva2:546041}.

\paragraph{Zero-Forcing Precoding}
The zero-forcing precoder is given by
\begin{align}\label{eq:ZF}
\mathbf{\tilde{W}}[\nu] = 
\alpha_\text{ZF} \estfreqchannel^\mathsf{H}[\nu] \big( \estfreqchannel[\nu] \estfreqchannel^\mathsf{H}[\nu] \big)^{-1} \hskip-1em,\hskip1em \quad \text{for \textsc{zf}}.
\end{align}
It nulls the interference $I_k$ at the cost of a lower gain $g_k$ compared to maximum-ratio precoding. For zero-forcing precoding and an i.i.d.\ Rayleigh fading channel, both with single-carrier and \OFDM transmission, the gain and interference terms are \cite{yang2013performance}
\begin{align*}
g_k = \sqrt{M - K}, \quad\quad
I_k = 0,\quad\quad\text{for \textsc{zf}}.
\end{align*}

\paragraph{Regularized Zero-Forcing Precoding}
Regularized zero-forcing precoding aims at maximizing the received \textsc{sinr} \eqref{eq:SINR}.  In the limit of an infinite number of antennas, the optimal linear precoder is given by \cite{Sanguinetti2014Optimal}
\begin{align}\label{eq:RZF}
\mathbf{\tilde{W}}[\nu] = 
\alpha_\text{RZF} \estfreqchannel^\mathsf{H}[\nu] \big( \estfreqchannel[\nu] \estfreqchannel^\mathsf{H}[\nu] {+} \rho \mathbf{I}_K \big)^{-1} \hskip-1em,\hskip1em \quad \text{for \textsc{rzf}},
\end{align}
where $\rho \in \mathbb{R}^+$ is a system parameter, which depends on the ratio $P T / N_0$ and on the path losses $\{\beta_k\}$ of the users.  The regularized zero-forcing precoder balances the interference suppression of zero-forcing and array gain of maximum-ratio precoding \cite{6832894} by changing the parameter $\rho$.  How to find the optimal parameter $\rho$ is described in \cite{Sanguinetti2014Optimal} and later in Section~\ref{sec:data_rate}.

The interference $I_k$ and gain $g_k$ of regularized zero-forcing depend on the parameter $\rho$ and no closed-form expression for them is known.  However, when the transmit power $P$ is low compared to the noise variance $N_0/T$, then a big $\rho$ is optimal and the interference and array gain are close to the ones of maximum-ratio precoding.  And when the transmit power relative the noise variance is high, a small $\rho$ is optimal and the interference and array gain are close to the ones of zero-forcing.

\subsection{Discrete-Time Constant-Envelope Precoding}\label{sec:DTCE}
The low-\PAR precoding scheme originally proposed in \cite{mohammed2012single} and extended in \cite{6451071, 6565529}, here called \emph{discrete-time constant-envelope} precoding, is briefly described in two sections, first for single-carrier transmission, then for \OFDM.

\subsubsection{Single-Carrier Transmission}
The discrete-time constant-envelope precoder produces transmit signals that have constant-envelope when viewed in discrete time, i.e.
\begin{align}
	|u_m[n]| = \frac{1}{\sqrt{M}}, \quad \forall n, m.
\end{align}
It does so by minimizing the difference between the received noise-free signal and the desired symbols under a fixed modulus constraint, $\{u_m[n]\} = $
\begin{align}\label{eq:MUI}
\argmin_{\{|u_m[n]| = M^{-1/2}\}} \sum_{n = 0}^{N - 1} \left\| \sum_{\ell = 0}^{L - 1} \mathbf{\hat{H}}[\ell] \mathbf{u}[n {-} \ell] - \sqrt{\gamma} \mathbf{s}[n] \right\|^2\!\!\!,
\end{align}
where $\gamma \in \mathbb{R}^+$ is a system parameter that is chosen to maximize the system performance.  Intuitively, a small $\gamma$ makes the interference $I_k$ small, but the gain $g_k$ small too.  On the other hand, a large $\gamma$ makes the array gain big but also makes it hard to produce the desired symbol at each user, which results in an excessive amount of interference.  In Section~\ref{sec:data_rate}, it will be shown how the parameter $\gamma$ is chosen such that the data rate is maximized.  

The optimization problem in \eqref{eq:MUI} can be approximately solved at a low computational complexity by cyclic optimization: minimizing the norm with respect to one $u_m[n]$ at a time, while keeping the other variables fixed.  Such a solver is not much heavier in terms of computations than the zero-forcing precoder \cite{6565529}.  

\subsubsection{OFDM Transmission}
\textsc{Ofdm} transmission in connection with discrete-time constant-envelope precoding can be done by using the same algorithm as for the single-carrier transmission.  Instead of precoding the symbols $\{\mathbf{s}[n]\}$ directly, the base station would precode their inverse Fourier transform
\begin{align}
	\mathbf{\tilde{s}}[n] \triangleq \frac{1}{\sqrt{N}} \sum_{\nu = 0}^{N - 1} e^{j 2 \pi \nu n / N} \mathbf{s}[\nu].
\end{align}

The convolution in \eqref{eq:MUI} should then be seen as a cyclic, i.e.\ the indices should be taken modulo $N$.

\subsection{Power Allocation among Users}
The power allocation $\{\xi_k\}$ between users has to be decided according to a chosen criterion, for example that all users shall be served with the same data rate.  This “egalitarian” criterion is used in this paper and is given by the max-min problem:
\begin{align}\label{eq:max_min_power_allocation_prob}
	\{\xi_k\} = \argmax_{\{\xi_k\}: \text{ eq.\eqref{eq:power_allocation_factor_constraint}}} \min_k \SINR_k,
\end{align}
where $\SINR_k$ is given in \eqref{eq:SINR}.  Note that out of all the terms in \eqref{eq:SINR}, apart from $\xi_k$ itself, only the clipping $c_k$, in-band distortion $D_k$ and the correlation $\rho_k$ might depend on $\xi_k$.  That $g_k$ and $I_k$ do not depend on $\{\xi_k\}$, can be seen from \eqref{eq:array_gain_step4} and \eqref{eq:app_MRP_interference_exp} in the Appendix.  Extensive simulations over Rayleigh fading channels indicate that only $D_k$ depends on the power allocation $\xi_k$ and that this dependence is linear.

To solve \eqref{eq:max_min_power_allocation_prob}, a first-order approximation of the dependence on $\xi_k$ is made.  The in-band distortion is assumed to be:
\begin{align}
	D_k = D' + \delta_k \xi_k D'',
\end{align}
where $D'$ and $D''$ are non-negative constants.  

For the $\{\xi_k\}$ that solve \eqref{eq:max_min_power_allocation_prob}, there is a common $\SINR$ such that $\SINR_k = \SINR$, for all $k$, because \eqref{eq:SINR} is an increasing function in $\xi_k$.  Rearranging \eqref{eq:SINR} gives the power allocation
\begin{align}\label{eq:optimal_power_allocations}
	\xi_k = \SINR \frac{P \beta_k ((I_k + E_k)|1+\rho_k|^2 + D') + \frac{N_0}{T}}{\delta_k P \beta_k (|g_k + c_k|^2 - D'' \SINR)}.
\end{align}
Because the power allocations sum to one \eqref{eq:power_allocation_factor_constraint},
\begin{align}\label{eq:common_SINR}
	\SINR \sum_{k=1}^{K} \frac{P \beta_k ((I_k + E_k)|1+\rho_k|^2 + D') + \frac{N_0}{T}}{\delta_k P \beta_k (|g_k + c_k|^2 - D'' \SINR)} = 1,
\end{align}
the common $\SINR$ can be found by solving this equation.  The optimal power allocations are thus given by \eqref{eq:optimal_power_allocations}, where $\SINR$  is the largest solution to \eqref{eq:common_SINR}.

Note that, if $D'' = 0$, \eqref{eq:common_SINR} can be solved explicitly, which gives an expression for $\SINR$ and the optimal power allocation
\begin{align}\label{eq:general_power_allocation}
	\xi_k = \frac{\frac{1}{\delta_k\beta_k |g_k + c_k|^2}(P \beta_k ((I_k + E_k)|1+\rho_k|^2 + D') + N_0/T)}{\sum_{k' = 1}^{K} \frac{P \beta_{k'} ((I_{k'} {+} E_{k'})|1+\rho_{k'}|^2 {+} D') {+} N_0/T}{\delta_{k'}\beta_{k'} |g_{k'} + c_{k'}|^2}}.
\end{align}

Specifically for maximum-ratio precoding when \eqref{eq:if_iid} holds, this power allocation becomes
\begin{align}\label{eq:power_allocation_MR}
	\xi_k^\text{MR} = 
	\frac{P \beta_k(|1+\rho_k|^2+D') + N_0/T}
	{\beta_k \delta_k \sum_{k' = 1}^{K} \frac{P \beta_{k'}(|1+\rho_{k'}|^2+D') + N_0/T}{\beta_{k'} \delta_{k'}}}, \quad \forall k,
\end{align}
and, for zero-forcing precoding, it becomes
\begin{align}\label{eq:power_allocation_ZF}
	\xi_k^\text{ZF} = 
	\frac{P \beta_k (E_k|1+\rho_k|^2 + D') + N_0/T}
	{\beta_k \delta_k \sum_{k' = 1}^{K} \frac{P \beta_{k'} (E_{k'}|1+\rho_{k'}|^2 + D') + N_0/T}{\beta_{k'} \delta_{k'}}}, \quad \forall k.
\end{align}
These two expressions \eqref{eq:power_allocation_MR} and \eqref{eq:power_allocation_ZF} are equivalent to the corresponding formulas in \cite{yang2014amacro} in the special case there is no amplifier distortion.  It should be noted that when the number of users is large, $K \gtrsim 30$, the term $D''$ is close to zero.

\subsection{Single-Carrier vs.\ OFDM Transmission}
In terms of the achievable data rate \eqref{eq:original_bound}, which has been proven tight when the channel hardens, single-carrier and \OFDM transmission are equivalent in massive \MIMO.  Due to channel hardening, all tones of the \OFDM transmission have equally good channels $\{\mathbf{\tilde{H}}[n]\}$, therefore the advantage of \OFDM---the possibility to do waterfilling across frequency---results in little gain.  This is summarized in the following Proposition and proven in the Appendix. % Appendix~\ref{app:equivalence_between_SC_and_OFDM}.
\begin{proposition}\label{pro:SC_OFDM_equ}
	If the same precoding scheme $f: \estfreqchannel[n] \mapsto \mathbf{\tilde{W}}[n]$ is used for all tones $n$, the rate in \eqref{eq:original_bound} is equal for single-carrier transmission \eqref{eq:tx_signal_SC} and for \OFDM \eqref{eq:tx_signal_OFDM}.
\end{proposition}

\begin{figure}
	\centering
	\includegraphics{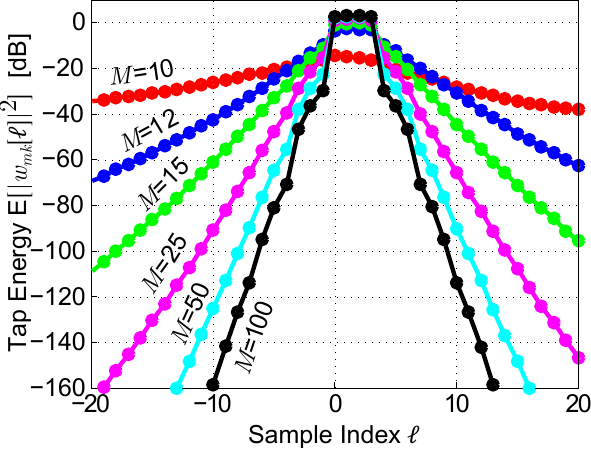}
	\caption{The normalized energy of the taps of the impulse response of the zero-forcing filter for single-carrier transmission over a frequency-selective 4\mbox{-}tap channel. The base station serves $K = 10$ users.  }
	\label{fig:ZF_filter_tap_profile}
\end{figure}

With regards to implementation, the two transmission methods differ.  While \OFDM requires a Fourier transform to be done by the users, single-carrier transmission does not.  While \OFDM causes a delay of at least $N$ symbols, since precoding and detection are done block by block, single-carrier transmission can be implemented for frequency-selective channels with short filters with much smaller delay.  Channel inversion with filters with few taps is only possible in massive \MIMO---in \textsc{siso} systems or small \MIMO systems, pre-equalization of a frequency-selective channel requires filters with a huge number of taps.  This can be seen in Figure~\ref{fig:ZF_filter_tap_profile}, which shows the power profile $\{\Expectation{|w_{mk}[\ell]|^2}, \mbox{$\ell = \ldots, -1, 0, 1, \ldots$}\}$ of the impulse response of the zero-forcing precoder for different numbers of base station antennas.  With few antennas, zero-forcing requires many filter taps, while, with massive \MIMO, it requires approximately the same number of filter taps as the number of channel taps.

Since the symbol period of \OFDM is longer than that of single-carrier transmission, $NT$ compared to $T$, \OFDM is less sensitive to synchronization errors in the sampling in \eqref{eq:rx_sampling}.  While a small time synchronization error, in the order of $T$, leads to a simple phase rotation in \OFDM, it would lead to difficult intersymbol interference in single-carrier transmission.  For small frequency synchronization errors, in the order $\frac{1}{NT}$, however, \OFDM suffers from intersymbol interference, while single-carrier transmission only experiences a simple phase rotation.

In some implementational aspects, the two transmission methods are similar though.  The computational complexities are roughly the same---the Fourier transform that single-carrier transmission spares the users from, has to be done by the base station instead, see \eqref{eq:SC_OFDM_relation}.  We stress that the signals of single-carrier and \OFDM transmission practically have the same \PAR in massive \MIMO, at least in an i.i.d.\ Rayleigh fading environment.  In Figure~\ref{fig:amp_dist}, there was only a small gap between the \PAR of single-carrier and \OFDM transmission and, when the number of channel taps is greater than $L > 4$, the gap is practically closed.

The operational differences between single-carrier and \OFDM transmission are summarized in Table~\ref{tab:comparison}.  

\section{Numerical Evaluations of Rate}
To estimate the power consumed by the amplifiers at different sum-rates and to compare different precoders to each other, the expectations in \eqref{eq:original_bound} that lack closed-form expressions were numerically evaluated for the system specified in Table~\ref{tab:sim_param}.  All continuous signals were modelled by $\kappa = 7$ times oversampled discrete-time signals.  Specifically the channel from antenna $m$ to user $k$ was assumed to be Rayleigh fading $h_{km}(\ell T/\kappa) \sim \mathcal{CN}(0, 1/(\kappa L))$, for $\ell = 0, \ldots, \kappa (L - 1)$ and i.i.d.\ across $k$, $m$ and $\ell$.  The users were assumed to be uniformly spread out over an annulus-shaped area, with inner radius $\varr$ and outer radius $100\varr$.  The path loss of user $k$ was then assumed to be
\begin{align}
	\beta_k = (\varr / \vard_k)^\alpha,
\end{align}
where $\vard_k$ is the distance between user $k$ and the base station, which is located in the middle of the annulus, and where $\alpha$ is the path loss exponent.

\begin{table}
	\centering
	\caption{System Parameters}
	\begin{tabular}{ll}
		\toprule
		number of tx-antennas & $M=100$                                            \\
		number of users       & $K=10$ and $50$                                    \\
		channel model         & $L=4$-tap Rayleigh fading                          \\
		pulse shape filter    & root-raised cosine, roll-off 0.22                  \\
		amplifier type        & class~B, see (\ref{eq:PA_eff}) and (\ref{eq:Rapp}) \\
		path loss exponent & $\alpha$ = 3.8 (typical urban scenario)\\ 
		\bottomrule
	\end{tabular}
	\label{tab:sim_param}
\end{table}

Further, it was assumed that the pilots used for channel estimation were $N_\text{p} = KL$ symbols long and sent with the same power $\rho_\text{p}$ from all users.  The power was chosen such that a signal sent from the cell edge, where the path loss is $\beta_\text{min} = 1/100^\alpha$, would be received at the base station with power \unit[0]{dB} above the noise variance, i.e.\ $\rho_\text{p} \beta_\text{min} = N_0/T$ and $\rho_\text{p} = 100^\alpha N_0/T$.

\begin{figure*}
	\centering
	\subfigure[The fraction of power lost due to clipping]{\label{fig:clipping_loss}\includegraphics{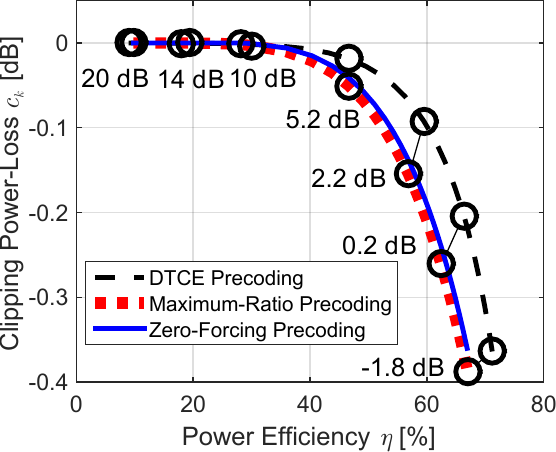}}
	\subfigure[The variance of the received in-band distortion]{\label{fig:NMSE}\includegraphics{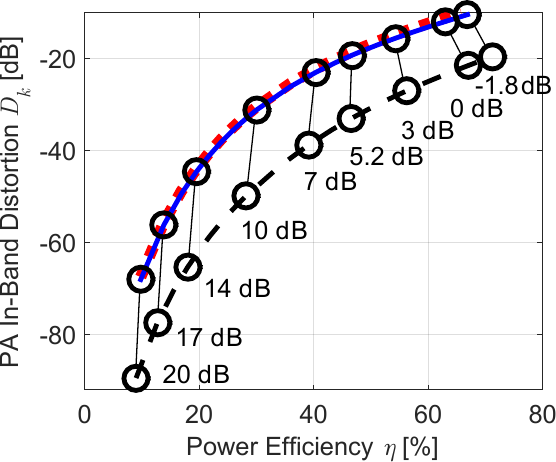}}
	\subfigure[The amount of out-of-band radiation]{\label{fig:ACLR}\includegraphics{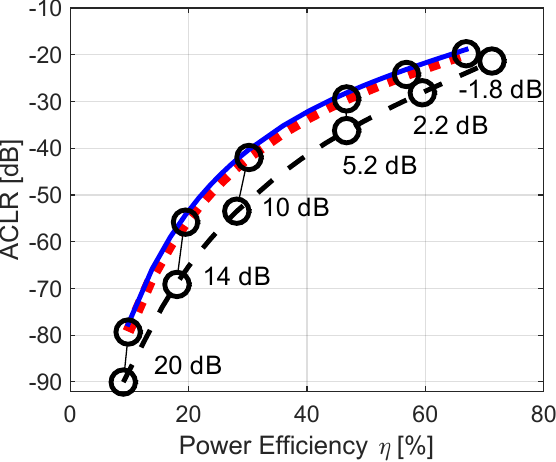}}
	\caption{Measurements on a Rapp-modeled ($p=2$) class~\textsc{b} amplifier for three signal types. The signals have been pulse shape filtered with a root-raised cosine, roll-off factor 0.22.  The encircled points correspond to some selected operating points of the amplifier, which are specified by the backoff from the 1-dB compression point.  It is assumed that all users are at the same distance to the base station, i.e.\ $\xi_k = 1/K$, for all $k$.\label{fig:NMSE_vs_eff}}
\end{figure*}

\subsection{Effects of Nonlinear Power Amplifiers}
The power amplifiers of the simulated system have been modeled by the Rapp model \cite{6457368}, where the phase distortion is neglected, so $\Phi(u) = 0$, $\forall u$, and the \textsc{am}-\textsc{am} conversion is given by
\begin{align}\label{eq:Rapp}
g(u) = A_\text{max} \frac{u / u_\text{max}} {(1 + (u / u_\text{max})^{2p})^{\frac{1}{2p}}},
\end{align}
where the parameter $p = 2$ approximates a typical moderate-cost solid-state power amplifier \cite{5357586}.  The parameter $A_\text{max}$ is the highest possible output amplitude and $u_\text{max} = A_\text{max}/g'(0)$ determines the slope of the asymptote that $g(u)$ approaches for small $u$.  

To ensure that the total radiated power is $P$, as is required by \eqref{eq:power_constraint}, the parameters are chosen as follows:
\begin{align}
u_\text{max} &= M^{-1/2}\\
A_\text{max} &= u_\text{max}\frac{\sqrt{P}}{\lambda_0},
\end{align}
where $\lambda _0$ is a correction factor to compensate for the power lost due to clipping, which is chosen such that \eqref{eq:power_constraint} holds.  Note that the correction factor is different for different signal types and backoffs.  

Massive \MIMO will require simple, inexpensive and power efficient amplifiers \cite{6690}.  The most basic class~\textsc{b} amplifiers have these properties \cite{raab2002power}, and could therefore potentially be suited for massive \MIMO.  The power efficiency of such an amplifier is given by \cite{6457368}
\begin{align}\label{eq:PA_eff}
\eta  = \frac{\pi }{4}\frac{\Expectation{g^2(|u_m(t)|)}}{A_\text{max}\Expectation{g(|u_m(t)|)}},
\end{align}
Note that $\eta  \leq \pi /4$, with equality only if the continuous-time input signal $u_m(t)$ has perfectly constant envelope and the amplifier operates in saturation.

The two phenomena of nonlinear amplification---in-band distortion and amplitude clipping---are studied by looking at the variance of the in-band distortion $\sigma_k^2 \triangleq \frac{D_k}{\xi_k |g_k|^2}$ and the clipped power $a_k \triangleq \smash{\frac{|g_k + c_k|^2}{|g_k|^2}}$ relative to the ideal amplitude.

The clipped power was computed together with the power efficiency of the amplifiers for several backoffs and averaged over many channel realizations for the system specified in Table~\ref{tab:sim_param}, in which all users are at the same distance to the base station and $\xi_k = 1/K$, for all $k$.  By treating the backoff as an intermediate variable, the clipped power can be given as a function of the efficiency, see Figure~\ref{fig:clipping_loss}.  It is noted that the power lost due to clipping is small (smaller than \unit[−0.4]{dB}) even when the amplifiers are operated close to saturation.

Similarly, the variance of the in-band distortion, Figure~\ref{fig:NMSE}, and the \ACLR, Figure~\ref{fig:ACLR}, were computed for several backoffs and averaged over different channel realizations.  It can be seen that the amount of energy radiated out-of-band is monotonically decreasing with the backoff.  A constraint on the \ACLR will therefore constrain the maximum efficiency that the amplifier can operate at.  Further, it is noted that the efficiency is not a simple function of the backoff, but it depends on the signal type.  We also note that, whereas the clipping power-loss is small at operating points with high efficiency, the in-band distortion (at least for the conventional precoders) and the out-of-band radiation are not.  The latter two phenomena will thus be the main factors to determine the operating point of the amplifiers.

Because of their similar amplitude distributions, all the linear precoding schemes (\textsc{mr, zf, rzf} precoding) result in similar curves in Figures~\ref{fig:clipping_loss}, \ref{fig:NMSE} and \ref{fig:ACLR}.  Therefore, only the results of single-carrier maximum-ratio and zero-forcing precoding are shown. The curves are identical to the ones of \OFDM transmission.

\begin{figure}
	\centering
	\includegraphics{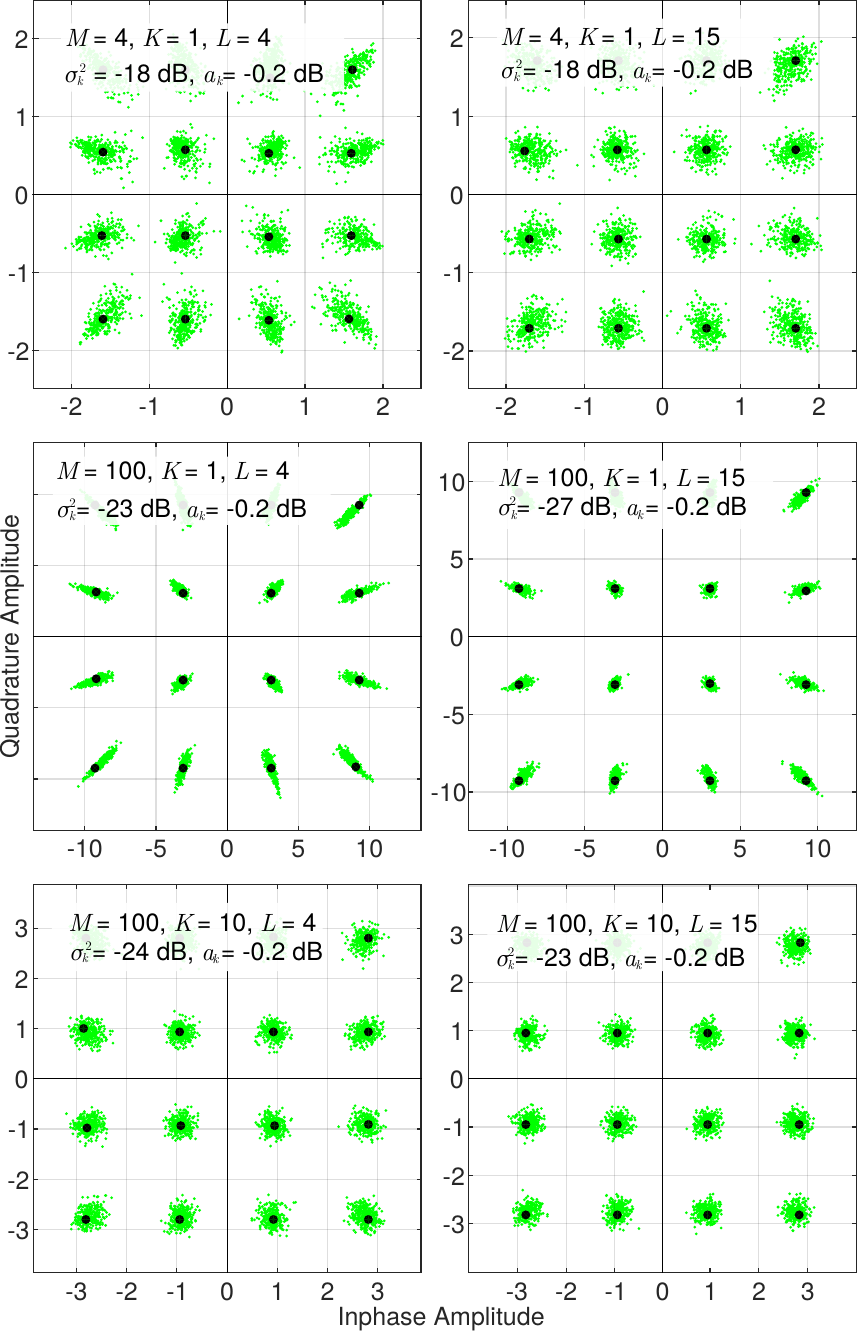}
	\caption{Received signal points without thermal noise when broadcasting 16-\textsc{qam} symbols with nonlinear amplification (\unit[1]{dB} backoff) over a \MIMO channel by single-carrier zero-forcing precoding for different number of users $K$, antennas $M$ and channel taps $L$.}
	\label{fig:RX_distortion}
\end{figure}

In Figure~\ref{fig:RX_distortion}, it can be seen how the effects of the nonlinear amplifiers change when the number of antennas, users and channel taps are changed in a single-carrier \MIMO system.  To make comparisons easy, all users in the system of Figure~\ref{fig:RX_distortion} have the same path loss and the amplifiers are backed off by \unit[1]{dB}, enough to see distinct clusters around each symbol point in all cases.  

When the number of users and channel taps are small, the distribution of the in-band distortion is different around different symbol points and the phase tends to be more accurately reproduced than the amplitude, resulting in oblong clouds around the outer symbol points.  It is a well-known phenomenon in \OFDM with a great number of subcarriers that the distribution of the in-band distortion is almost circularly symmetric Gaussian and identically distributed around each symbol point, which means that the distortion can be regarded as uncorrelated additive noise \cite{costa1999impact}.  In multiuser \MIMO, a similar effect is observed---when the number of users and channel taps are great, the noise is almost circularly symmetric Gaussian and identically distributed around each symbol point for single-carrier transmission too.  This is intuitive, since the precoded transmit signals are the sums of many independent symbols and the receive signals are the sums of many different transmit signals.  The in-band distortion therefore gets mixed up and its distribution becomes symmetric and identical around all symbol points, just as is the case in \OFDM.

In Figure~\ref{fig:RX_distortion}, the variance of the in-band distortion seems to roughly follow the scaling law predicted by \cite{bjornson2014massive}, which says that the distortion variance relative to the signal power scales as $\ordo(\sqrt{K / M})$.  Although the in-band distortion seems to disappear with increasing number of antennas, the amplitude reduction due to clipping does not (in Figure~\ref{fig:RX_distortion}, it remains at \unit[−0.2]{dB} for all system setups), which was not observed in \cite{bjornson2014massive}.  However, it can be argued that the clipping power-loss is small and only needs to be considered when the amplifiers are operated close to saturation.

\subsection{Data Rate and Power Consumption}\label{sec:data_rate}
In this section the power $P_\text{cons}$ that the base station amplifiers consume is estimated.  Even if the discrete-time constant-envelope precoder outperformed the other precoders in the comparisons in Figure~\ref{fig:NMSE_vs_eff}, in the end, it is in terms of consumed power the precoders should be compared.

The rate $R_k(P, \theta)$ in \eqref{eq:original_bound} is a function of the transmit power $P = \eta P_\text{cons}$, and therefore a function of the operating point of the power amplifiers, which are parametrized by their efficiency $\eta$.  In the case of discrete-time constant-envelope precoding, the rate is also a function of the parameter $\theta = \gamma$.  And in the case of regularized zero-forcing, it is a function of the parameter $\theta = \rho$.  For a given out-of-band radiation requirement, specified by a maximum \textsc{aclr} level $\ACLR_\text{max}$, the sum-rate of the system is thus given by
\begin{align}\label{eq:rate}
	R(P_\text{cons}) = \max_{\eta, \theta} \sum_{k=1}^{K} R_k(\eta P_\text{cons}, \theta),
\end{align}
where the maximization is over all $\theta \in \mathbb{R}^+$ and over all operating points $\eta \in [0, \eta_\text{max}]$, where $\eta_\text{max}$ is the highest operating point that still results in an \textsc{aclr} below $\ACLR_\text{max}$.  If the \ACLR is not constrained, $\eta_\text{max}$ is taken to be the maximum possible efficiency of the given amplifiers and signal type.

\begin{figure}
\centering
\includegraphics{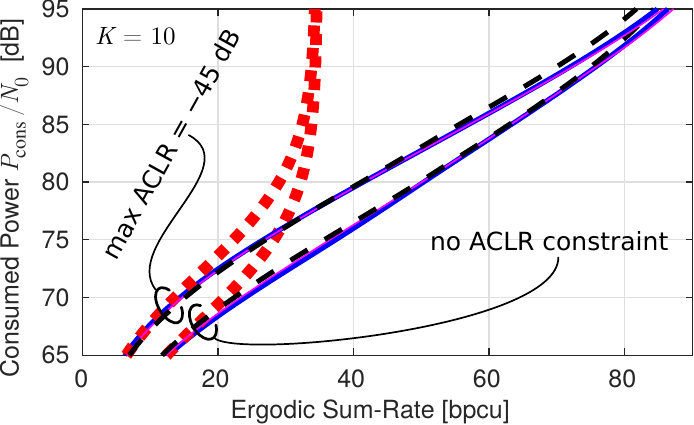}
\null\hspace{0.75em}\includegraphics{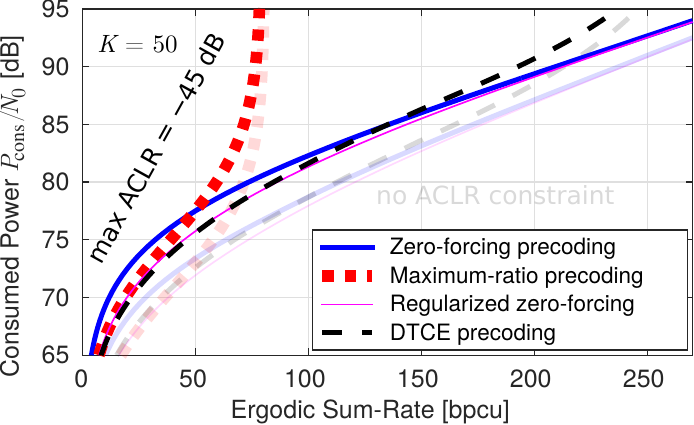}
\caption{The estimated consumed power of a base station with $M=100$ antennas required to serve $K = 10$ (above) and $K = 50$ (below) users with a certain rate over a frequency-selective channel with $L = 4$ taps with and without a constraint on the \ACLR.}
\label{fig:rate_vs_cons_power}
\end{figure}

The relation between consumed power and the average sum-rate of the system that is shown in Figure~\ref{fig:rate_vs_cons_power} has been obtained by computing \eqref{eq:rate} for many different user distributions $\{\beta_k\}$ and taking an average.  Both the cases (i) when the out-of-band radiation is constrained by requiring the \ACLR to be below \unit[−45]{dB}, which is the \ACLR requirement in \textsc{lte} \cite{3GPP_TS36.141_LTE_BS_testing}, and (ii) when it is not constrained are considered.

It can be seen that maximum-ratio precoding works well for low rate requirements but is limited by interference to below a certain maximum rate. Because the \SINR in \eqref{eq:SINR} contains distortion that scales with the radiated power, all precoders have a vertical asymptote, above which the rate cannot be increased.  Except for discrete-time constant-envelope precoding, whose curve starts to bend away upwards in the plot for 50 users, this vertical asymptote lies outside the scale and cannot be seen for the other precoders however.  Since the array gain $|g_k|^2$ is smaller for discrete-time constant-envelope precoding than for zero-forcing and regularized zero-forcing, its vertical asymptote is located at a lower rate than the asymptote of zero-forcing and regularized zero-forcing precoding.

Further, it can be seen that regularized zero-forcing and zero-forcing perform equally well when the number of users is small.  Regularized zero-forcing has an advantage over zero-forcing when the number of users is big though, because of its ability to balance the resulting array gain and the amount of interuser interference received by the users.  

The low-\PAR precoding scheme, discrete-time constant-envelope precoding, seems to consume roughly the same amount of power as the conventional precoding schemes, both when the \ACLR is constrained and when it is not, at least for low to medium rate requirements.  At very high rates, the optimal linear precoder has an advantage over discrete-time constant-envelope precoding---because the vertical asymptote of discrete-time constant-envelope precoding is at a lower rate than it is for the optimal linear precoder.

\begin{figure}
\centering
\includegraphics{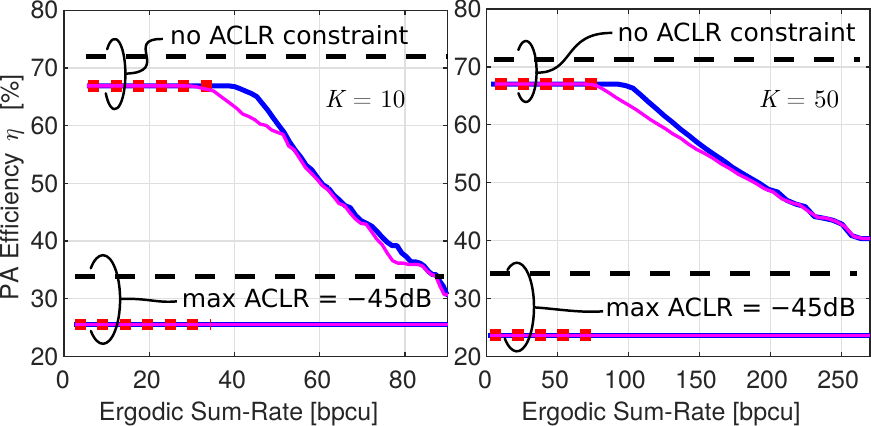}
\caption{The power efficiency of the amplifiers at the optimal operating point for different sum-rate requirements.  The legend in Figure~\ref{fig:rate_vs_cons_power} also applies here.}
\label{fig:eff_vs_rate}
\end{figure}

The value of $\eta$ that corresponds to the optimal operating point of the amplifiers is shown in Figure~\ref{fig:eff_vs_rate}.  When there is no constraint on the out-of-band radiation, it is optimal to operate the amplifiers in saturation, where the power efficiency is high, for low rate requirements.  For higher rate requirements, the amplifiers should be backed off to lower the in-band distortion for the conventional precoding schemes.  The amplifiers of the low-\PAR precoding scheme, however, continue to operate close to saturation also at high rates.  When the \ACLR is constrained to below \unit[−45]{dB}, the optimal efficiency of the amplifiers coincides with $\eta_\text{max}$ (the highest permissible operating point), i.e.\ \unit[34]{\%} for discrete-time constant-envelope precoding and \unit[27]{\%} for maximum-ratio and zero-forcing precoding, over the whole range of rates investigated, both when serving 10 and 50 users.  This corresponds to a backoff of \unit[8]{dB} and \unit[11]{dB} respectively.

\section{Conclusions}\label{sec:conclusions}
We have compared four different multiuser \MIMO precoders: maximum-ratio, zero-forcing, regularized zero-forcing and discrete-time constant-envelope precoding.  They can be used for single-carrier transmission and for \textsc{ofdm} transmission.  The different precoders and transmission methods are summarized in Table~\ref{tab:comparison}.

In massive \MIMO, there is little operational difference between \textsc{ofdm} and single-carrier transmission in terms of complexity, and in terms of \PAR.  It also turns out that single-carrier and \OFDM transmission have the same performance in terms of data rate.  Additionally, massive \MIMO allows for time-domain channel inversion to be done with a short filter, for which the number of taps is of the same order of magnitude as the number of channel taps.  This makes single-carrier transmission easy to implement and allows for running precoding with little delays.  Since \textsc{ofdm} requires the users to be equipped with an additional \textsc{fft}, this would suggest that single-carrier transmission should be considered in massive \MIMO systems.  

A massive \MIMO system with centralized baseband processing also allows for low-\PAR precoding, which increases the power efficiency of the amplifiers but requires more radiated power to compensate for the lowered array gain compared to conventional precoders.  For the simplistic power amplifier model used, simulations have shown that the amplifiers of the base station consume the same amount of power when using low-\PAR precoding as when using the optimal conventional precoder.  Since low-\PAR transmit signals are more hardware-friendly and could enable cheaper and simpler base station designs, this would suggest that a low-\PAR precoding scheme that also pre-equalizes the channel and suppresses interference (such as discrete-time constant-envelope precoding) should be used in massive \MIMO base stations with centralized baseband processing.

Furthermore, in massive \MIMO, simulations have shown that the power efficiency of the amplifiers at the optimal operating point is limited by the out-of-band radiation requirement and that in-band distortion caused by nonlinear amplifiers has two parts: one clipping part that decreases the amplitude of the received signal and one part that can be seen as additive uncorrelated noise, which drowns in the thermal noise in representative cases.  The amplitude lost due to clipping is small even when the amplifiers are operated close to saturation.

\newcommand{\specialcell}[2][t]{%
  \begin{tabular}[#1]{@{}P{\linewidth}@{}}#2\end{tabular}}

\begin{sidewaystable}
    \centering
\footnotesize
\caption{Precoding Schemes and Transmission Methods for Massive \textsc{Mimo}}
\label{tab:comparison}
\begin{tabular}{lm{0.8cm}P{0.2\linewidth}P{0.24\linewidth}P{0.24\linewidth}}
\toprule
&  &  & Single-Carrier\tikzmark{s2}{}  & \tikzmark{f2}{}\textsc{Ofdm}  \\ 
\midrule\rule{0pt}{2.2ex}
&& general &  \specialcell{Simple matched-filter detection\\Running precoding and detection, little delay\\Sensitive to time synchronization errors} & \specialcell{Matched-filter detection in frequency domain (requires \textsc{fft} at the users)\\Block-wise precoding and detection\\Sensitive to frequency synchronization errors} \\ 
\midrule%
\raisebox{2ex}{\multirow{3}{*}{\rotatebox[origin=c]{90}{Linear Precoders -- high \PAR}}}&
\textsc{Mr}  & \specialcell{Maximizes array gain\\Enables local precoding; distributed arrays\\Low complexity\\Performs well at low data rates} & \specialcell{Has residual interference\\$L$\mbox{-}tap \textsc{fir} filter\\\textsc{Sinr} as in \eqref{eq:SINR}, with $|g_k|^2 = M, I_k = \delta_k$} & \specialcell{Has residual interference\\\textsc{Sinr} as in \eqref{eq:SINR}, with $|g_k|^2 = M, I_k = \delta_k$} \\ 
\cline{2-5}
&\textsc{Zf}  & \specialcell{Nulls interuser and intersymbol interference\\Requires inversion of $K{\times}K$\mbox{-}matrices} & \specialcell{${\sim }L$\mbox{-}tap \textsc{fir} filter\\\textsc{Sinr} as in \eqref{eq:SINR}, with $|g_k|^2 {=} M{-}K$, $I_k {=} 0$} & \specialcell{\textsc{Sinr} as in \eqref{eq:SINR}, with $|g_k|^2 {=} M{-}K$, $I_k {=} 0$} \\ 
\cline{2-5}
&\textsc{R-Zf}  & \specialcell{Linear precoder that maximizes \textsc{sinr}\\Optimization over parameter $\rho$ needed\\Performance-wise similar to \textsc{zf} for small $K$, but better for big $K$\\Requires inversion of $K{\times}K$\mbox{-}matrices} & \specialcell{${\sim }L$\mbox{-}tap \textsc{fir} filter\\\textsc{Sinr} as in \eqref{eq:SINR}, with $|g_k|^2 \in [M{-}K, M]$, $I_k\in[0,\delta_k]$} & \specialcell{\textsc{Sinr} as in \eqref{eq:SINR}, with $|g_k|^2 \in [M{-}K, M]$, $I_k\in[0,\delta_k]$} \\ 
\midrule
&\textsc{Dtce}  & \specialcell{Low \PAR\\Nonlinear precoder\\Optimization over parameter $\gamma$ needed}& Filter with delay ${\sim }L$ &  \\ 
\bottomrule
 \end{tabular} 
     \renewcommand{\ArcDistance}{0.2cm}
          \begin{tikzpicture}[overlay,remember picture]
              \draw[
                      <->, thick, distance=\ArcDistance,
                      shorten <=\ShortenBegin, shorten >=\ShortenEnd,
                      out=0, in=180, black
                  ] 
                      ($(s2.east) + (1em, -1ex)$) to 
                      ($(f2.north) + (-1em, -1ex)$);
              \node [] at ($(s2.north)!0.5!(f2.north) + (+2em, -0.67ex) + (0,\ArcVector)$) {\tiny \parbox[t][\height][c]{4cm}{equivalent when waterfilling and\\ joint sequence detection is not done.}};
          \end{tikzpicture}
\end{sidewaystable}

\appendices 
\appendix[Proof of Proposition~\ref{pro:SC_OFDM_equ}]
\label{app:equivalence_between_SC_and_OFDM}
We show that single-carrier and \OFDM transmission, both with a cyclic prefix added in front of each transmission block, result in the same achievable rate \eqref{eq:original_bound}.  To do that, the effects of the amplifiers are neglected.  However, it is reasonable to assume that the in-band distortion caused by the amplifiers affects the two transmission schemes in the same way given that the amplitude distributions and \PAR of the two transmission schemes are the same.  The data rate of single-carrier and \OFDM are the same if the array gains $|g_k^\text{SC}|^2 = |g_k^\text{OFDM}|^2$ and interference variances $I_k^\text{SC} = I_k^\text{OFDM}$ are the same.

We start by expanding the array gain \eqref{eq:channel_est} for single-carrier transmission:
\begin{align}
	&|g_k^\text{SC}|^2 = \frac{1}{\xi_k^2 \delta_k} \Big|\mathsf{E}\Big[s_k^*[n] \sum_{m}\sum_{\ell}\hat{h}_{km}[\ell] u_m[n - \ell]\Big]\Big|^2\\
	&{=} \frac{1}{\xi_k^2 \delta_k} \! \Big| \mathsf{E}\Big[\! s_k^*[n] \!\sum_{m,\ell}\! \hat{h}_{km}[\ell] \!\sum_{k',\ell'}\! w_{mk}[\ell'] s_{k'}\hskip-0.1em[n{-}\ell{-}\ell'] \Big] \! \Big|^{\mathrlap{2}}
\end{align}
Since different symbols are uncorrelated and since they have energy $\xi_k$, only terms for which $k' = k$ and $\ell = -\ell'$ will remain, so
\begin{align}\label{eq:array_gain_step2}
	|g_k^\text{SC}|^2 = \frac{1}{\delta_k} \Big| \mathsf{E} \Big[ \sum_{m} \sum_{\ell} \hat{h}_{km}[\ell]w_{mk}[-\ell]\Big] \Big|^2.
\end{align}
This is a cyclic convolution evaluated in $0$, which can be computed in the frequency domain followed by an inverse transform:
\begin{align}
	|g_k^\text{SC}|^2 = \frac{1}{\delta_k} \Big| \sum_{m} \frac{1}{N} \sum_{n} \mathsf{E}\Big[ \hat{\tilde{h}}_{km}[n]\tilde{w}_{mk}[n] \Big] \Big|^2.
\end{align}
If the same precoding scheme $f$ has been used for all frequencies, i.e.\ $f: \estfreqchannel[n] \mapsto \mathbf{\tilde{W}}[n], \forall n$, then all the terms in the inner sum are the same:
\begin{align}\label{eq:array_gain_step4}
	|g_k^\text{SC}|^2 = \frac{1}{\delta_k}  \Big|\sum_{m} \mathsf{E}\Big[ \hat{\tilde{h}}_{km}[0]\tilde{w}_{mk}[0] \Big] \Big|^2,
\end{align}
which is also the array gain $|g_k^\text{OFDM}|^2$ of \OFDM.

We now study the interference \eqref{eq:interference} for single-carrier transmission, by using \eqref{eq:array_gain_step2}:
\begin{align}
	&I_k^\text{SC} = \mathsf{E}\Big[ \Big| \sum_{m} \sum_{\ell} \hat{h}_{km}[\ell] u_m[n - \ell] - g_k s_k[n] \Big|^2 \Big]\\
	&= \mathsf{E}\bigg[ \Big| \sum_m \sum_{\ell} \hat{h}_{km}[\ell] \sum_{k'} \sum_{\ell'} w_{mk'}[\ell'] s_{k'}[n{-}\ell{-}\ell']\notag\\
	&\hskip4.5em - s_k[n] \mathsf{E}\Big[ \sum_m \sum_{\ell} \hat{h}_{km}[\ell] w_{mk}[-\ell] \Big] \Big|^2 \bigg]\\
	&= \mathsf{E} \bigg[ \! \Big|\! s_k[n] \Big( \!\sum_{\mathclap{m, \ell}}\! \hat{h}_{km}[\ell] w_{mk}[-\ell] {-} \mathsf{E}\Big[\! \sum_{\mathclap{m, \ell}}\! \hat{h}_{km}[\ell] w_{mk}[-\ell] \Big] \!\Big) \notag\\
	&\hskip2.5em + \underset{\mathclap{(k',n') \neq (k, 0)}}{\sum\sum} s_{k'}[n'] \sum_m\sum_{\ell} \hat{h}_{km}[\ell] w_{mk'}[n'-\ell] \Big|^2\bigg]
\end{align}
Since different symbols are uncorrelated and since they have energy $\xi_k$, the square is expanded into the following.
\begin{align}
	I_k^\text{SC} = \sum_{k'}\sum_{n'} \xi_{k'} \mathsf{E}\Big[ \Big| \sum_m \sum_{\ell} \hat{h}_{km}[\ell] w_{mk'}[n' - \ell] \Big|^2\Big] \notag\\
	- \xi_k \Big| \mathsf{E}\Big[ \sum_m \sum_{\ell} \hat{h}_{km}[\ell] w_{mk}[-\ell] \Big]\Big|^2
\end{align}
The two terms are cyclic convolutions in $n'$ and $0$ respectively and can be computed in the frequency domain.  Along the same line of reasoning as in \eqref{eq:array_gain_step4}, the interference variance is given by $I_k^\text{SC} =$
\begin{align}\label{eq:app_MRP_interference_exp}
	\hskip-\parindent \sum_{k'}\! \xi_{k'} \mathsf{E}\Big[ \big| \!\sum_m\! \hat{\tilde{h}}_{km}[0] \tilde{w}_{mk'}[0] \big|^2\Big] {-} \xi_k\Big| \mathsf{E}\Big[ \!\sum_m\! \hat{\tilde{h}}_{km}[0] \tilde{w}_{mk}[0] \!\Big]\! \Big|^2\!\!,
\end{align}
which is precisely the interference variance $I_k^\text{OFDM}$ of \OFDM at tone $0$, or at any other tone.

That the rate \eqref{eq:original_bound} is the same for single-carrier and \OFDM transmission was proven, in a different way, for the special case maximum-ratio precoding in \cite{diva2:546041}.

\bibliographystyle{IEEEtran}
\bibliography{bib_hela_namn,bibliografi}
\end{document}